\documentclass[twocolumn,preprintnumbers,superscriptaddress,amsmath,amssymb,floatfix]{revtex4}
\usepackage{graphicx}
\usepackage{dcolumn}
\usepackage{bm}
\usepackage{psfrag}
\usepackage{epsfig}
\usepackage{amsmath}
\usepackage{amssymb}
\usepackage{mathbbol}
\usepackage{xcolor}
\usepackage{epstopdf,txfonts}
\usepackage{subcaption} 

\newcommand{\beq}{\begin{equation}}
\newcommand{\eeq}{\end{equation}}

\begin{document}

\title{Beat-to-Beat Fetal Heart Rate Analysis Using Portable Medical Device and Wavelet Transformation Technique}

\author{Maria Farahi}
\affiliation{Sana Meditech S.L., Barcelona, Spain}
\affiliation{Center of Research in Biomedical Engineering. Universitat Politècnica de Catalunya, BarcelonaTech, Barcelona, Spain}

\author{Al\'{i}cia Casals}
\affiliation{Center of Research in Biomedical Engineering. Universitat Politècnica de Catalunya, BarcelonaTech, Barcelona, Spain}

\author{Omid Sarrafzadeh}
\affiliation{Sana Meditech S.L., Barcelona, Spain}

\author{Yasaman Zamani}
\affiliation{Sana Meditech S.L., Barcelona, Spain}

\author{Hooran Ahmadi}
\affiliation{Sana Meditech S.L., Barcelona, Spain}

\author{Naeimeh Behbood}
\affiliation{Sana Meditech S.L., Barcelona, Spain}

\author{Hessam Habibian}
\affiliation{Sana Meditech S.L., Barcelona, Spain}

\date{\today}

\begin{abstract}
A beat-to-beat Tele-fetal Monitoring and comparison with clinical data are studied with a wavelet transformation approach. Tele-fetal monitoring is a big progress toward a wearable medical device for a pregnant woman capable of obtaining prenatal care at home. We apply a wavelet transformation algorithm for fetal cardiac monitoring using a portable fetal Doppler medical device. Choosing an appropriate mother wavelet, 85 different mother wavelets are investigated. The efficiency of the proposed method is evaluated using two data sets including public and clinical. From publicly available data on PhysioBank, and simultaneous clinical measurement, we prove that the comparison between obtained fetal heart rate by the algorithm and the baselines yields a promising accuracy beyond 95\%. Finally, we conclude that the proposed algorithm would be a robust technique for any similar tele-fetal monitoring approach.
\end{abstract}

\maketitle

\section{Introduction}
Gestation can end with a live birth, a spontaneous miscarriage, an induced abortion, or a stillbirth \cite{ Abman2011}. Prenatal care by the mother and constant monitoring of the gestational period are key elements in improving birth outcomes \cite{Reichman2010}. Clinically, various medical devices have been introduced to monitor the Fetal Heart Rate (FHR) during pregnancy. Approaches like cardiotocography (CTG) \cite{grivell2015antenatal}, fetal magnetocardiography (fMCG) \cite{peters2001monitoring}, fetal electrocardiography (fECG) \cite{nassit2015non}, and fetal scalp electrocardiography (fsECG) \cite{organ1968scalp} are examples of clinical techniques that can be applied for fHR monitoring.
 
First of all, fetal scalp electrocardiography in which electrodes are applied on the fetal scalp. It captures signals with a high Signal to Noise Ratio (SNR). However, it is invasive and increases the risk of infections. In addition, it can be used only during the delivery. And, it needs a skilled specialist for installation \cite{organ1968scalp}.

 Secondly, among non-invasive devices, fMCG is a method which is well known for its high SNR in data capturing. It uses SQUID (for Superconducting QUantum Interference Device) sensors to record the magnetic field of the fetal heart from maternal abdomen. But it is expensive and needs a shielded room with an expert specialist to apply it on the mother's abdomen\cite{peters2001monitoring}.

Next non-invasive technique, known as fECG, is a cheap technology which enables users to perform continuous monitoring. This method not only suffers from low SNR, but also its electrodes should be placed on the mother's abdomen by a skilled specialist \cite{nassit2015non}. 

 Finally, CTG is a standard non-invasive clinical method. It is a well-known technique which uses Doppler ultrasound sensors to monitor both fHR and uterus contractions. It is highly accurate, and requires less skills to operate \cite{nageotte2015fetal, euliano2013monitoring}, but it is highly sensitive to the fetal movements \cite{alnuaimi2019fetal}.

Tele-Fetal Monitoring (TFM) is an approach which makes a pregnant woman capable of obtaining prenatal care. It can decrease the risk of pregnancies with hypertensive disorders \cite{lanssens2018impact}. Moreover, it has illustrated profits in other high-risk pregnancies like those with gestational diabetes. Also, it can be beneficial in having access to rural pregnant women who are far from modern hospitals \cite{van2018ehealth}. 

In the past decade, many efforts have been dedicated to find a suitable way for distant fHR extraction. As an example, fHS analysis is considered as a non-invasive method. It is simple to apply and it is low-cost. However, fHS signals are profoundly corrupted by noise since they are recorded at the maternal abdominal surface. There are various sources of noise in fHS signals including fetal movements, contractions of mother's uterus, maternal digestive sounds, sensor movements, ambient sounds, maternal respiratory and maternal heart sound \cite{strazza2018pcg}.
 
The basic principle behind the fHS processing is that the heart’s mechanical activity is accompanied by the generation of various characteristic sounds. These sounds are associated with changes in the speed of blood flow, as well as with the opening and closing of heart valves \cite{tang2016phonocardiogram}. Dia et al. estimated adult heart rate from phonocardiograph (PCG) signals \cite{dia2019heart}. They applied a non-negative matrix factorization approach on the spectrogram of the taken signals. They evaluated their work by considering synchronous ECG and PCG signals. Samieinasab et al. \cite{samieinasab2015fetal}, used a single-channel denoising framework to reduce the noise of fetal PCGs. Then, similar to \cite{dia2019heart}, they utilized non-negative matrix factorization method to decompose fPCGs in time-frequency domain. In addition, Khandoker et al. proposed a four-channel fPCG system and evaluated it using fECG data \cite{khandoker2018validation}. They used 10-minutes of clinical data for evaluation. Reported results were P\textless 0.01 in cross correlation analysis and \textless 5 \% agreement in Bland-Altman plot.

Due to the fact that the most challenging step in fHR extraction from fetal heart sound signals is denoising, this paper presents a denoising algorithm based on wavelet transformation technique. Furthermore, the aim of the current paper is to introduce an algorithm capable of performing fHR extraction from fetal heartbeat sound in order to facilitate tele-fetal monitoring for pregnancies.

Throughout the investigation in this paper, we perform a comprehensive investigation among different mother wavelets to find the most accurate fHR extraction method from experimental fHS signals. We consider different 85 mother wavelets including Daubechies (order from 1 to 45), Symlets (order from 1 to 20),  Coiflet (order from 1 to 5), and Bioorthogonal (order from 1.1 to 6.8). We investigate mentioned wavelet families in different levels of decomposition from 1 to 12. In total 1020 investigations is performed. Here the methodology is applied over a tele-monitoring ultrasound device, i.e., we  use a pocket-size portable doppler device to measure the fHR. We use a wired protocol and save the data through a mobile application developed in-house and based on the proposed method. Using the portable pocket-size device and the proposed method, we aim to embed the fHR algorithm in future at the mobile App level for the end-user (mothers) as well as a cloud-based solution for gynecologist to monitor pregnancies remotely.


In the following, in section 2, the procedure of clinical measurement, data acquisition and applied methods for fHR extraction is described. We explain in detail the pre-processing and wavelet filtering ans processes established for the fHR evaluation. In section 3, we analyze, validate and discuss the results. Finally, the the results are discussed and concluded in sections 4 and 5, respectively.

\begin{figure*}[!h]
\centering
{\includegraphics[width=0.7\linewidth]{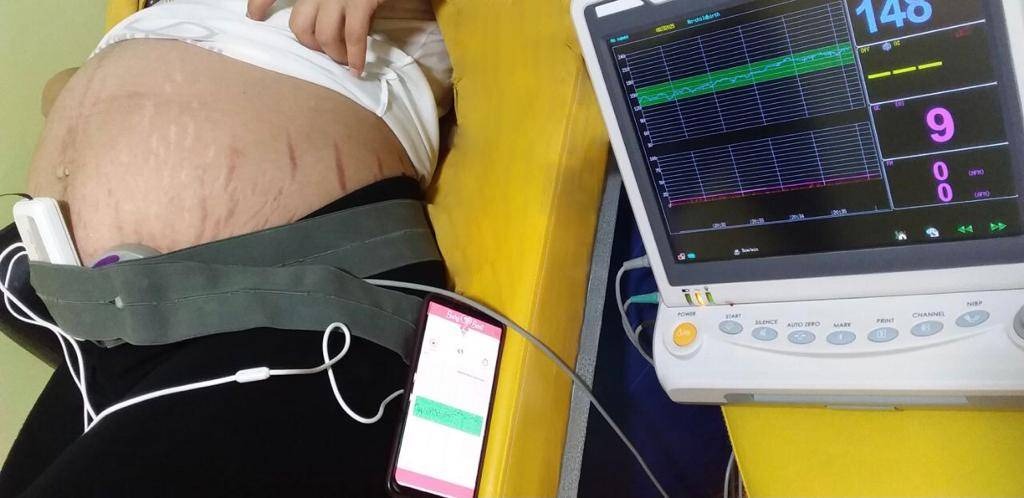}%
}

\caption{Clinical simultaneous measurement of Baby Sound A pocket fetal doppler from Contec Medical \cite{ContecMedicalCo} and CTG clinical device used from Bionet brand (FC-1400).}
\label{fig:DataAcquisition}
\end{figure*}

\section{Clinical measurement, data acquisition and methods}
The clinical data collection is based on a simultaneous measurement of a portable (pocket-size) ultrasound doppler device and a CTG clinical device as depicted in  Fig. \ref{fig:DataAcquisition}.
The portable device used is Baby Sound A pocket fetal doppler from Contec Medical \cite{ContecMedicalCo} which is under a distribution brand of Baby Heart Beat from Sana Meditech company\footnote{www.babyheartbeat.eu}. The portable device is certified under medical CE and FDA approval. 
On the other hand the simultaneous measurement is performed by a clinical device for validation and to be used as a gold standard (baseline). The clinical device used is a from Bionet brand with model FC-1400.

\begin{figure*}[!t]
\centering
\subfloat[]{\includegraphics[width=0.5\linewidth]{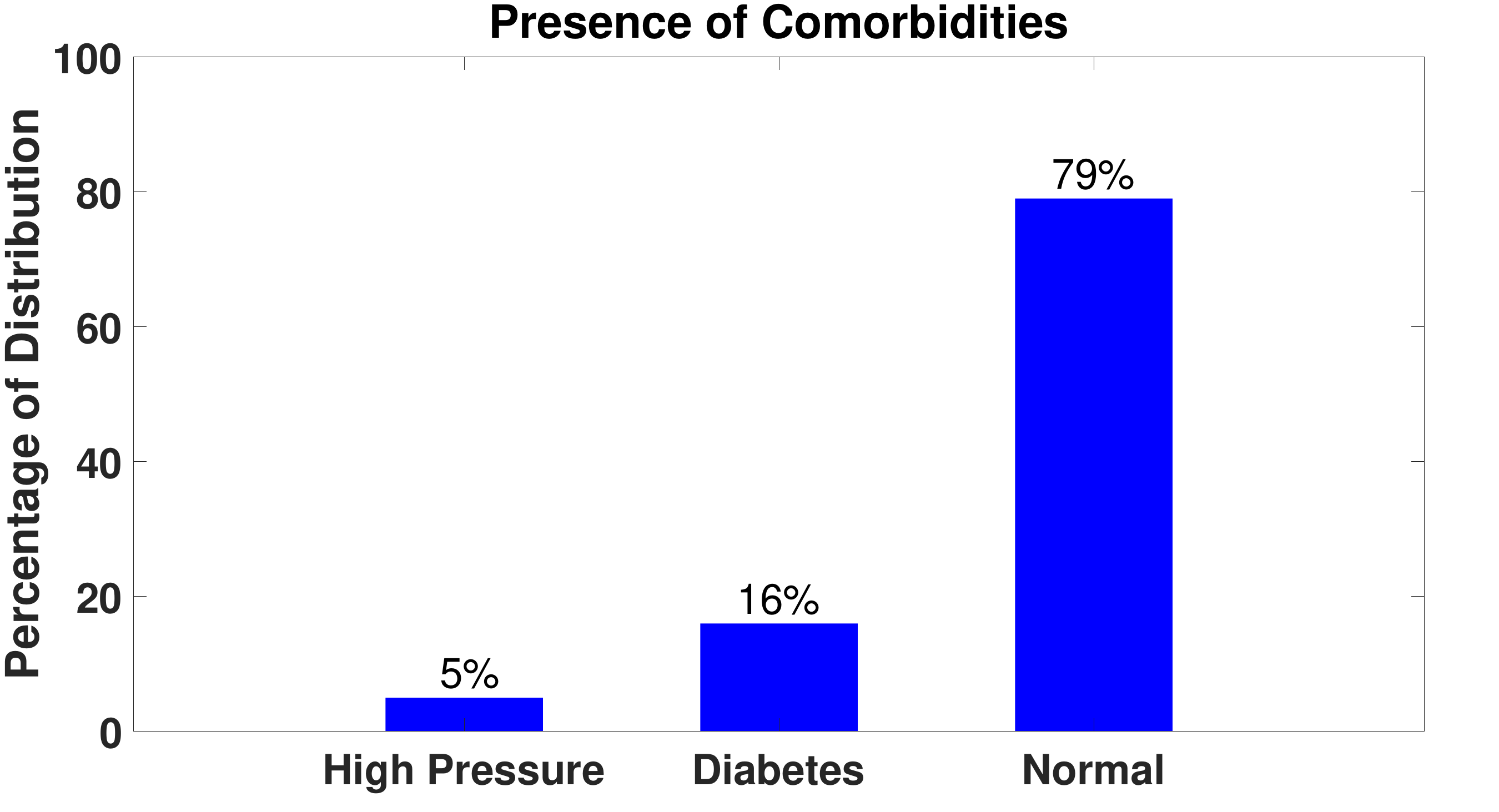}}
\hfil
\subfloat[]{\includegraphics[width=0.5\linewidth]{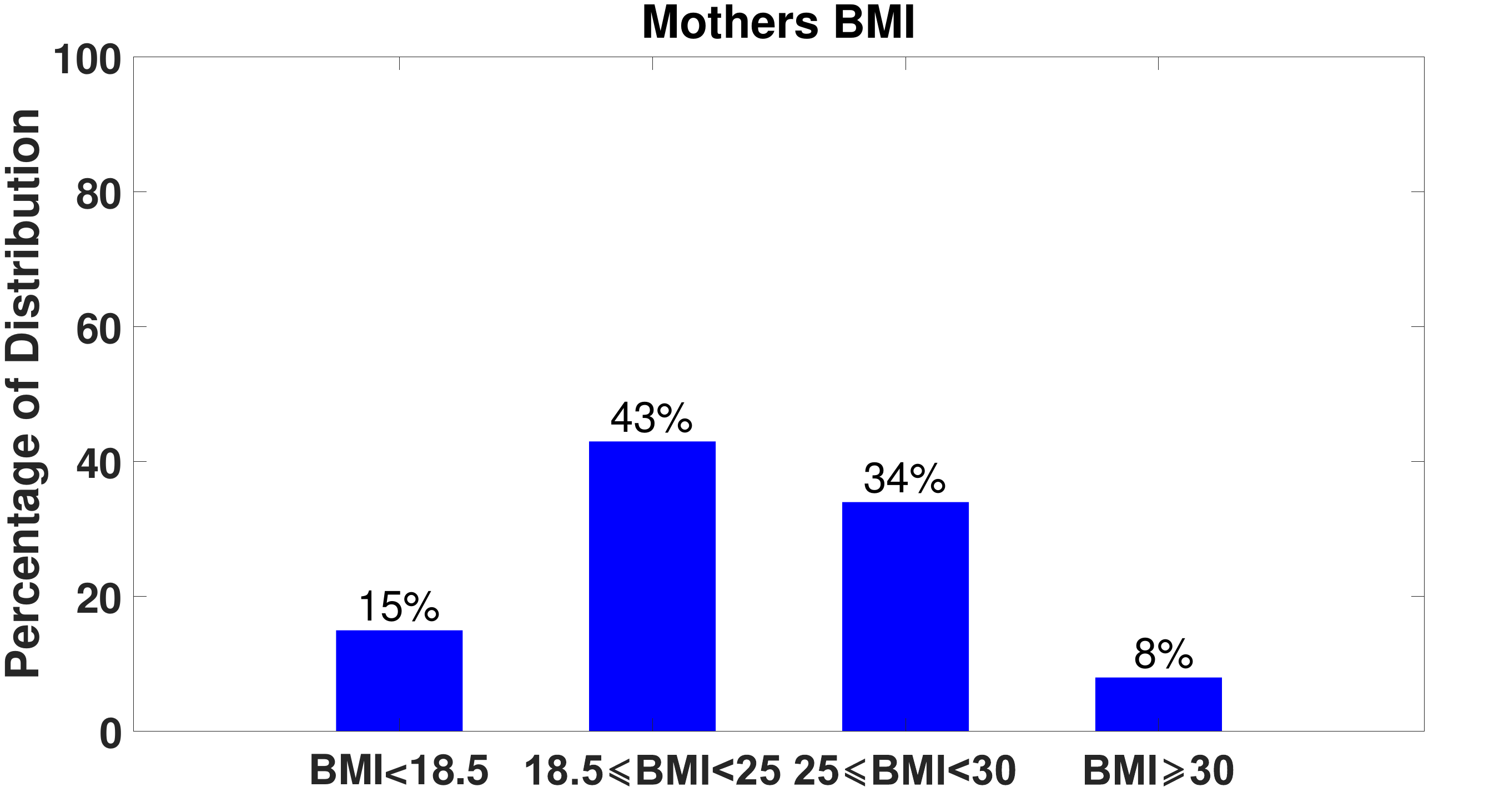}}
\hfil
\subfloat[]{\includegraphics[width=0.5\linewidth]{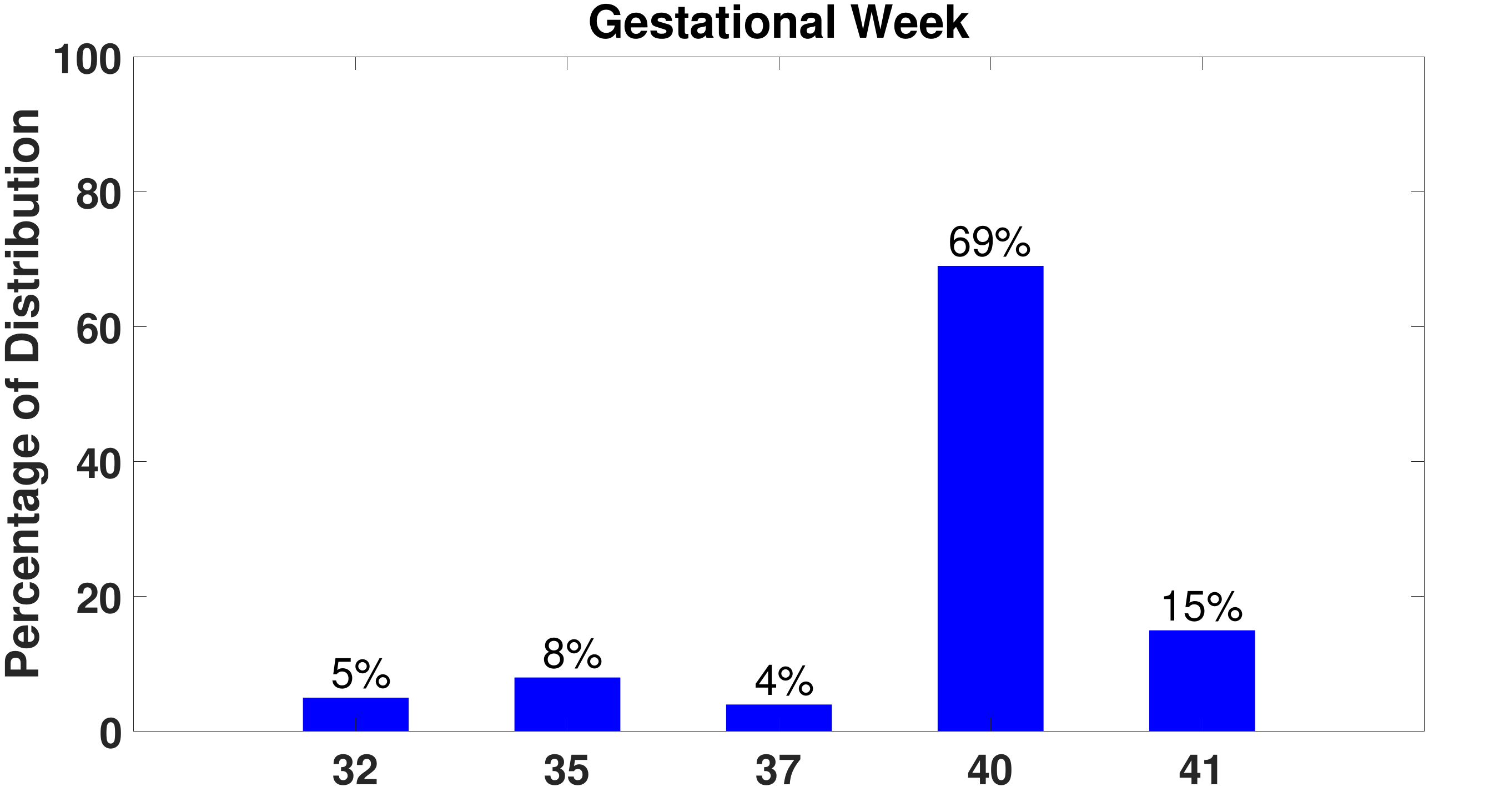}}
\hfil
\subfloat[]{\includegraphics[width=0.5\linewidth]{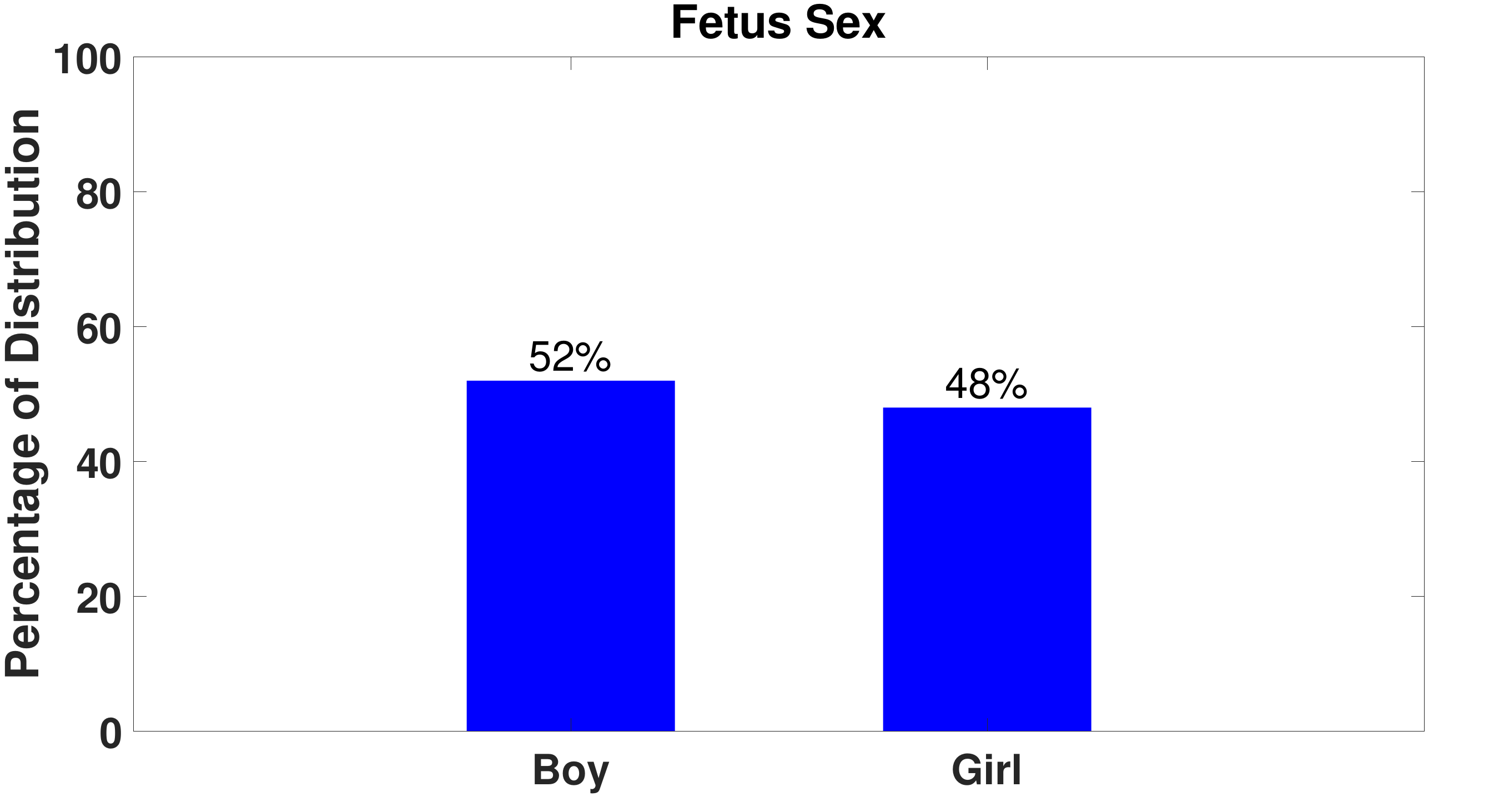}}
\caption{Statistical distribution of clinical data of pregnant women. The distributions of (a) Normal vs high risk pregnancies, (b) BMI of pregnant moms before pregnancy, (c) gestational pregnancy week, (d) gender of the fetus, are depicted.  }
\label{fig:StatisticalDistribution}
\end{figure*}

Simultaneous measurement would allow us to validate the data collected by the portable device and the software build to process the fHR, as an alternative to the clinical CTG device. Total number of 131 samples is acquired from a various pregnant women with ages range between 19 to 43 years old (average age of 29 years old). The length of each sample considered throughout this paper is always more than 1 minute and sampling frequency rate of 8 kHz. 
In Fig. \ref{fig:StatisticalDistribution} statistical summary of the captured clinical data distribution is provided. Among all target patients, 5\% are high pressure pregnant women while around 21\% are patients with pregnancy diabetes and the rest are in a healthy (normal) situation. Moreover, we report the pregnant women's Body Mass Index (BMI) before pregnancy, as depicted in the Fig. \ref{fig:StatisticalDistribution} the minimum BMI is 17.44 while the maximum is 32.83 with an average and standard deviation of 24.11 and 4.0, respectively. Finally, 52\% of the fetuses are targeted to be boys, and 48\% are supposed to be girls. The pregnancy gestational age distribution is also reported with a maximum abundance of 40 weeks (about 69\%).
As a summary we report a brief summary of clinical data acquired in Table \ref{tab: studyPopulationCriteria} for the sake of completeness. Clinical data is recorded for rigorous analysis of our algorithm and continuous improvement during the development process. The Table \ref{tab: studyPopulationCriteria} illustrates the eligible criteria of our study population.

\begin{table}[!htt]
\caption{Brief summary of clinical data acquired population study and eligible criteria}
\label{tab: studyPopulationCriteria}
\centering
\begin{tabular}{|c|c|c|c|c|c|c|c|c|c|}
\hline
\multicolumn{10}{|c|}{Eligible Criteria}\\
\hline
\multicolumn{2}{|c|}{\scriptsize Age (year)} & \multicolumn{2}{|c|}{\scriptsize Type of pregnancy} & \multicolumn{2}{|c|}{\scriptsize Gestation (week)} & \multicolumn{2}{|c|}{\scriptsize Body Mass Index} & \multicolumn{2}{|c|}{\scriptsize Anomaly} \\
\hline
\multicolumn{2}{|c|}{\scriptsize from 18 to 50} & \multicolumn{2}{|c|}{\scriptsize singleton} & \multicolumn{2}{|c|}{\scriptsize greater than 32} & \multicolumn{2}{|c|}{\scriptsize from 15 to 45} & \multicolumn{2}{|c|}{\scriptsize allowed} \\
\hline
\end{tabular}
\end{table}

The portable fetal monitoring device of Baby Heart Beat (Baby Sound A pocket fetal doppler) has been chosen because of its high quality, design and audio precision with respect to other similar portable devices. The device has an AUX port which provides a possibility to export fHS data directly to an external storage like an smart cell phone. The exported fHS signal is recorded by a dedicated mobile App developed internally by Sana Meditech company. In the following, we propose a new method consisting of adaptive band pass filtering and wavelet transformation in order to extract fHR from recorded fHS signals.

\begin{figure}
\centering
\includegraphics[width=0.9\linewidth]{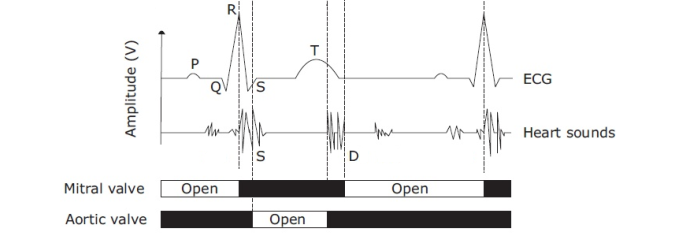}%
\hfil
\caption{A schematic representation of a single heartbeat cycle from an ultrasound probe. Temporal difference between heart sound and ECG signals are shown \cite{schmidt2007detection}.}
\label{fig:S1S1}
\end{figure}

\begin{figure}[!ht]
\centering
\includegraphics[width=0.9\linewidth]{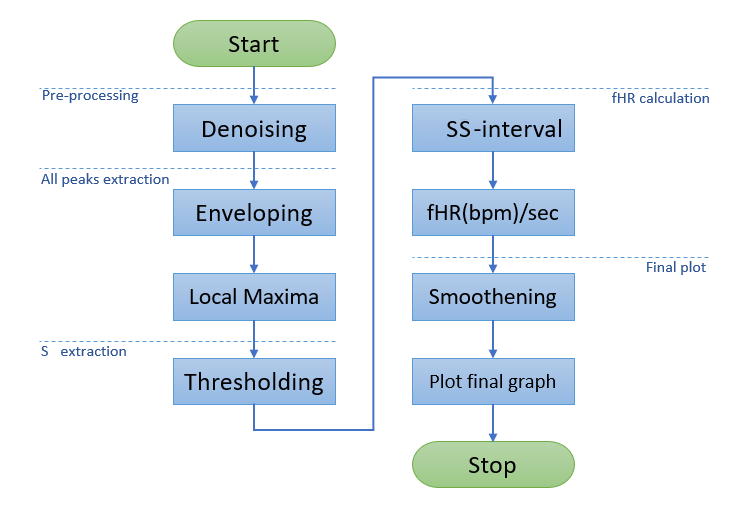}
\caption{Flowchart of the steps for our algorithm development for the fHR evaluation.}
\label{fig:Drawing1}
\end{figure}

The Fetal heart sound signal collected by an ultrasound doppler device and from the mother's abdomen would have a shape like Fig. \ref{fig:S1S1}. In this figure, S shows the position of systole while D is representative of Diastole. In order to extract fHR, the distance between two systole should be calculated. 
For our investigation, we have followed a process for our algorithm development which is shown in the Flowchart of Fig. \ref{fig:Drawing1} and depicts overall signal processing steps for fHR extraction from fHS signals. As seen, the algorithm starts with a pre-processing. In later steps, those peaks which are related to systole are identified, and ultimately the fHR is extracted. Finally, a smoothing step is performed. 
In the following sections we describe the major steps shown in the flowchart.

\subsection{Pre-processing: pre-development algorithm with a simulated data}

For the pre-development and testing of our algorithm, we initially use a simulated fetal heartbeat sound data set provided by Cesarelli \cite{cesarelli2012simulation} which is publicly available in physionet \cite{simulatedFetalPCGs}. It contains 37 signals with a duration of about 8 minutes each, and sampling frequency is 1 kHz.
As pointed out by the authors in \cite{cesarelli2012simulation} in order to simulate environmental noises, this data set signals are provoked by different SNR values from -26.7dB to -4.4dB. For more information about this data set, we refer the reader to reference \cite{simulatedFetalPCGs}. 

Fetal heart sound signals are heavily corrupted by noise since they are recorded at the maternal abdominal surface. Fetal movements, contractions of mother's uterus, mother's abdominal sounds, sensor movements, ambient sounds, maternal respiratory and heart sound are the various sources of noise \cite{strazza2018pcg}.
 In this study, noise reduction is performed based on the use of wavelet transformion technique. 

\subsubsection{Wavelet Transformation} 
The basic idea behind Wavelet Transformation (WT) is to define a new basis function which can be enlarged or compressed to capture both low frequency and high frequency component of the signal. Here we use WT to denoise and clean our signal. Mathematically WT is a time-frequency processing method and its definition for an input signal $x(t)$ is given Eq. \ref{eq:WaT}:

\begin{equation}
\label{eq:WaT}
WT_{x(a,b)}=\int x(t) \Psi^*_{a,b}(t)dt \hspace{0.5cm} ; a \neq 0
\end{equation}
\noindent where the basic function, $\Psi_{a,b}(t)$, is featured by scale and time-shift parameters ($a$ and $b$, respectively) as Eq.\ref{eq:EqPsi}:
\begin{equation}
\label{eq:EqPsi}
\Psi_{a,b}(t)= \frac{1}{\surd a} \psi (\frac{t-b}{a})
\end{equation}

$\Psi_{a,b}(t)$ is also been used for signal decomposition. The main challenge in using WT for denoising is to choose the optimum mother wavelet for the given tasks. In order to systematically decide on which mother wavelet is suitable for a specific purpose, main properties such as vanishing moments, size of support, regularity, orthogonality, bio-orthogonality, energy, symmetry, and the ability of implementing on discrete signals are investigated \cite{yang2019parameterised, rhif2019wavelet}. In Table \ref{tab:motherWavelets} we summarize 14 different families of mother wavelets that we have investigated in this paper. 
Over the next section we describe in detail the procedure of choosing the best chosen mother wavelet accordingly for the fHS denoising.

\begin{table*}[!t]
  \caption{A Summary of Different Mother Wavelets' Properties}
  \begin{tabular}{p{0.19\linewidth}p{0.03\linewidth}p{0.03\linewidth}p{0.05\linewidth}p{0.03\linewidth}p{0.03\linewidth}p{0.05\linewidth}p{0.03\linewidth}p{0.03\linewidth}p{0.03\linewidth}p{0.09\linewidth}}
   \hline
  \textbf{\scriptsize{\rotatebox[origin=c]{-90}{Mother Wavelet}}} & \textbf{\scriptsize{\rotatebox[origin=c]{-90} {Compact Support}}}& \textbf{\scriptsize{\rotatebox[origin=c]{-90}{Vanishing Moment}}}&\textbf{\scriptsize{\rotatebox[origin=c]{-90}{Regularity}}} &\textbf{\scriptsize{\rotatebox[origin=c]{-90}{Orthogonal}}} & \textbf{\scriptsize{\rotatebox[origin=c]{-90} {Bio Orthogonal}}} & \textbf{\scriptsize {\rotatebox[origin=c]{-90}{Symmetric}}} & \textbf{\scriptsize{\rotatebox[origin=c]{-90}{Energy}}}& \textbf{\scriptsize{\rotatebox[origin=c]{-90}{Discrete Wavelet}}}&\textbf{\scriptsize{\rotatebox[origin=c]{-90}{Continues Wavelet}}}&\textbf{\scriptsize{\rotatebox[origin=c]{-90}{References}}}\\
  
   \hline
   {\scriptsize HAAR} & \checkmark & 1 & \texttimes &\checkmark& \checkmark & \checkmark  & \checkmark &\checkmark & \checkmark & \cite{Daubechies1992} \\
    \hline
   {\scriptsize Daubechies} & \checkmark & {\scriptsize N} & {\scriptsize 0.2N} & \checkmark & \checkmark  & \texttimes & \checkmark&\checkmark &\checkmark & \cite{Daubechies1992} \\
     \hline
  {\scriptsize Symlets} &\checkmark & {\scriptsize N} & - &\checkmark & \checkmark & {\scriptsize near} & \checkmark &\checkmark&\checkmark & \cite{Daubechies1992}\\
    \hline
  {\scriptsize Coiflets} & \checkmark & {\scriptsize 2N} & - &\checkmark & \checkmark & {\scriptsize near} & \checkmark &\checkmark&\checkmark &\cite{Daubechies1992}\\
    
     \hline
  {\scriptsize Biorthogonal} & \checkmark & {\scriptsize Nr} & {\scriptsize Nr-1} & \texttimes & \checkmark & \checkmark & \checkmark &\checkmark & \checkmark & \cite{Daubechies1992}\\
    \hline
  {\scriptsize Fejer-Korovkin} & \checkmark & {\scriptsize N} & - & \checkmark & \texttimes & \texttimes & \checkmark &\checkmark& \checkmark & \cite{nielsen2001construction,chen2020comparing}\\
    
     \hline
  {\scriptsize Meyers} & \texttimes & {\scriptsize N} & {\scriptsize inf} &\checkmark& \checkmark &\checkmark& \checkmark & *&\checkmark & \cite{Daubechies1992}\\
    
    \hline
  {\scriptsize Gaussian} & \texttimes & - & - & \texttimes &\texttimes & **& \texttimes&\texttimes&\checkmark &\cite{mallat1999wavelet}\\
    \hline
  {\scriptsize Mexican hat} & \texttimes & {\scriptsize 2} & - & \texttimes& \texttimes &\checkmark & \texttimes & \texttimes&\checkmark & \cite{Daubechies1992, Leigh2013}\\
     \hline 
  {\scriptsize Morlet} & \texttimes & - & - &\texttimes & \texttimes & \checkmark& \texttimes & \texttimes &\checkmark & \cite{Daubechies1992}\\
     \hline 
  {\scriptsize Complex Gaussian}  & \texttimes & - & - & \texttimes & \texttimes & ** & \texttimes & \texttimes&\checkmark &\cite{mallat1999wavelet}\\
    \hline
  {\scriptsize Shannon}  & \texttimes & {\scriptsize N} & - & \texttimes& \texttimes& \checkmark&\texttimes & \texttimes&\checkmark & \cite{teolis1998computational}\\
    \hline
    {\scriptsize Freq. B-Spline} & \texttimes& - & - & \texttimes& \texttimes& \checkmark & \texttimes &\texttimes&\checkmark & \cite{teolis1998computational}\\
     \hline
  {\scriptsize Complex Morlet} &\texttimes & - & - &\texttimes& \texttimes&\checkmark &\texttimes &\texttimes&\checkmark & \cite{teolis1998computational}\\
      \hline
  \end{tabular}\\
  
  \footnotesize {* It is possible but without fast WT.\\
  ** They are symmetric if their order is even.}
  \label{tab:motherWavelets}
\end{table*}

\subsubsection{Mother wavelet Selection}
As explained, the procedure we follow is that the selection of the most suitable mother wavelet is performed by looking at each mother wavelet's properties (see table \ref{tab:motherWavelets}). Therefore firstly we discard mother wavelets which are far from our goal which is denoising of fHS signals. For instance, we need a mother wavelet that preserves the energy of the signal after decomposition (like orthogonal and bio-orthogonal). As a result, we cross out non-orthogonal mother wavelets including Gaussian, Mexican hat, Morlet, Complex Gaussian, Shannon, Freq B-Spline and Complex Morlet. Similarly, the chosen wavelet family should be able to offer the possibility of performing the discrete wavelet transformion. Since the Meyer family can not do fast WT for discrete data, we exclude it from our further investigation.

Relatively complex mother wavelets, with a minimum number of vanishing moments are needed. This will allow to represent more complex functions with fewer wavelet coefficients \cite{Omari2014}. Considering the mentioned specifications, mother wavelets of Daubechies, Symlets, Coiflets and Biorthogonal are chosen. In this paper, all mother wavelets of $db1$, $db2$, $\dots$, $ db45$, $sym1$, $sym2$ ,$ \dots$, $sym20$, $coif1$, $coif2$, $\dots$, $coif5$, and $bior1.1$, $bior1.2$, $\dots$, $bior6.8$
are investigated.

In order to find the most proper mother wavelet among others, two factors are important to be taken into account: \textbf{energy} and \textbf{entropy}. Energy clarifies how much a signal and a mother wavelet are similar to each other. The energy of a detail signal at each resolution level, $j$ is:
\begin{equation}
\label{eq:Energy}
E_j= \sum_{j=1}^{J}|C_j(k)|^2 
\end{equation}
where $C_j(k)$ is wavelet coefficients in level $j$. In consequence, the total energy can be obtained by:

\begin{equation}
\label{eq:energyTotal}
E_{tot}= \sum_{j}\sum_{k}|C_j(k)|^2|= \sum_{j=1} E_j
\end{equation}

Relative wavelet energy will assist to choose an effective mother wavelet. It can detect the degree of similarities between different segments of a signal \cite{salwani2005relative} and is defined as Eq. \ref{eq:RWE}. Entropy shows the effect of the mother wavelet on the accuracy of reconstruction. It illustrates how much data will be missed by a chosen mother wavelet. The definition of entropy is presented in Eq. \ref{eq:entropy}, where $p_{j}$ is the energy probability distribution of the wavelet coefficients defined in Eq. \ref{eq:RWE}.

\begin{tabular}{p{3.5cm}p{5.5cm}}
  \begin{equation}
  \label{eq:RWE}
  p_j=\frac{E_j}{E_{tot}}
  \end{equation}
 &
 \begin{equation} \label{eq:entropy}
  H(j) = -\sum_{j=1}^{J} p_{j}^2 log(p_{j}^2)
\end{equation}
\end{tabular}

By Dividing relative wavelet energy by entropy (see Eq. \ref{EEReq}) we obtain a ratio that clearly indicates which mother wavelet mostly resembles the original signal. The mother wavelet we are mostly interested in, will be the one that obtains a higher value of this ratio, meaning that the similarities between the wavelet and the original signal are greater than the non-conserved information between them.
\begin{equation}
\label{EEReq}
RWEER(j) = \frac{E(j)}{H(j)}
\end{equation}
where RWEER is a representative of Relative Wavelet Energy to Entropy Ratio.

\begin{figure*}[!t]
  \centering
  \includegraphics[width=\linewidth]{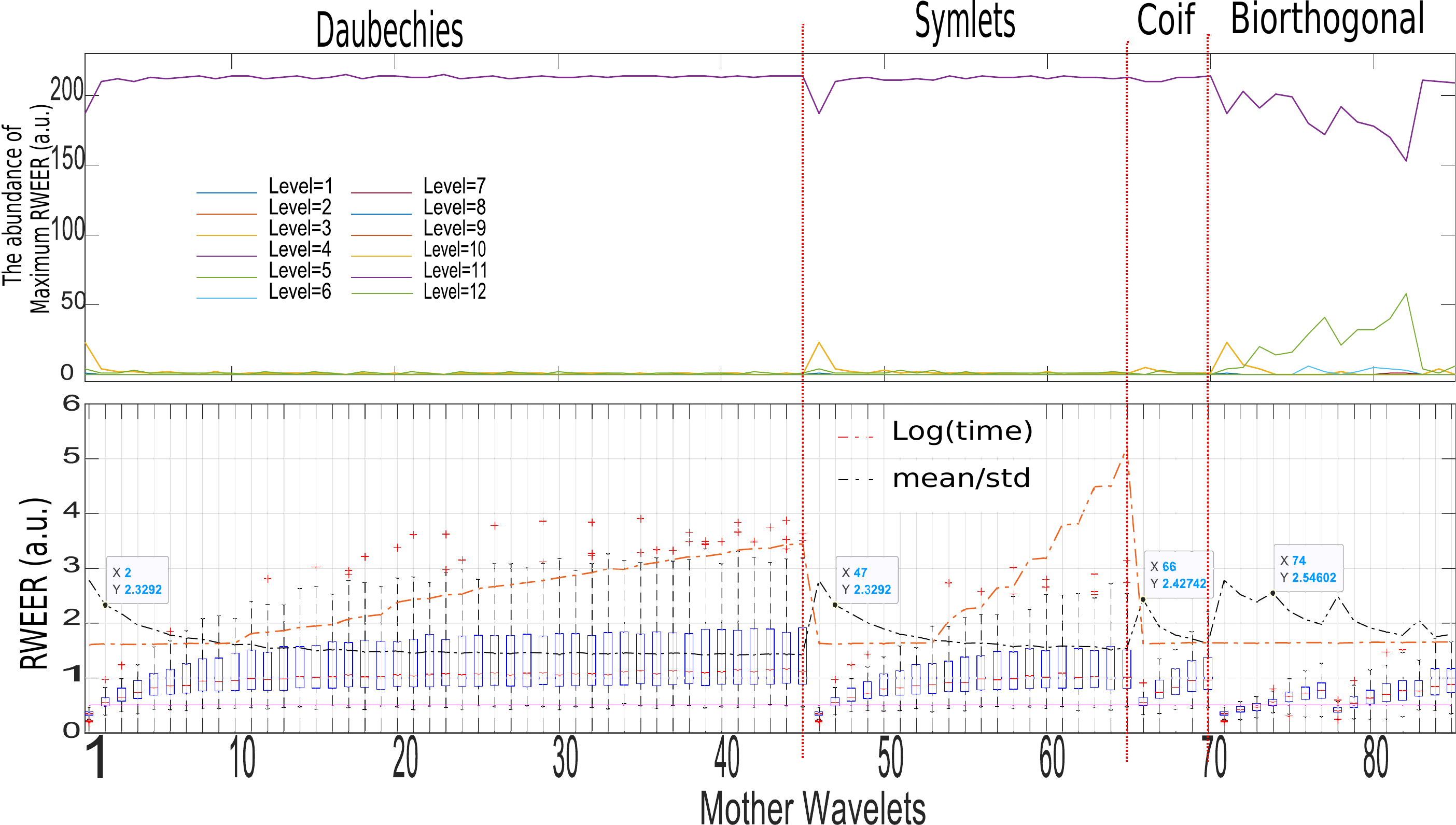}
  \caption{Distribution plot of RWEER for different 85 mother wavelets including $db$ 1, $db$ 2, \dots, $db$ 45 ,$sym$ 1, $sym$ 2, \dots, $sym$ 20, $coif$ 1, $coif$ 2, \dots, $coif$ 5, $bior$ 1.1, $bior$ 1.3, \dots, and $bior$ 6.8. First row: The effect of wavelet decomposition in different levels on the abundance of maximum RWEER. It shows that level $j=4$ works better than others. Second row: The investigation of RWEER for 215 sections with different mother wavelets on level $j=4$, red dash-dot graph interprets the proposed algorithm's execution time for each mother wavelet, the black dash-dot graph represents the value of $\frac{mean}{std}$ for shown box-plot. Outliers are drawn in red plus.}
  \label{fig:MWI}
\end{figure*}

\begin{figure*}
\centering
\subfloat[]{\includegraphics[width=0.5\linewidth]{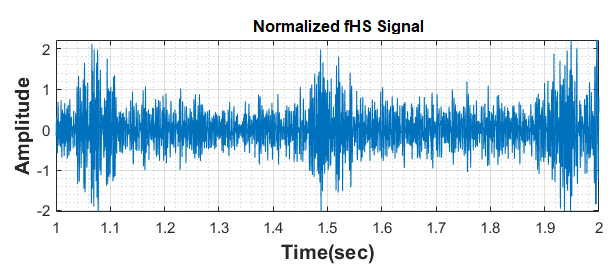}%
\label{fig:wavthresh1}}
\hfil
\subfloat[]{\includegraphics[width=0.5\linewidth]{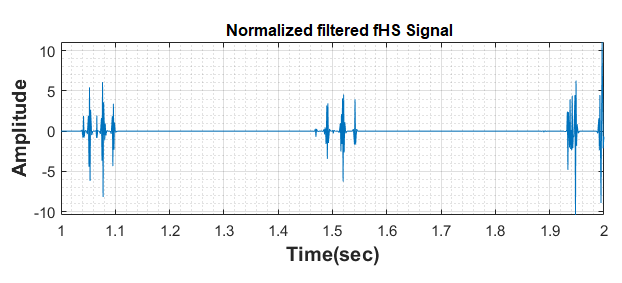}%
\label{fig:wavthresh2}}
\hfil
\subfloat[]{\includegraphics[width=0.5\linewidth]{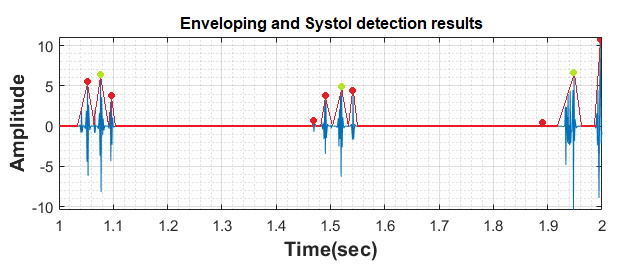}%
\label{fig:wavthresh3}}

\caption{A sample of an original fHS signal before and after filtering: (a) Original signal, (b) Denoised signal using the proposed method, and (c) Extracted S-peaks. Red graph shows the obtained envelop and red bullets represent all local maxima. Green bullets illustrate the position of selected S.}
\label{fig:wavthresh}
\end{figure*}

\begin{figure*}[!t]
\centering
\includegraphics[width=\linewidth]{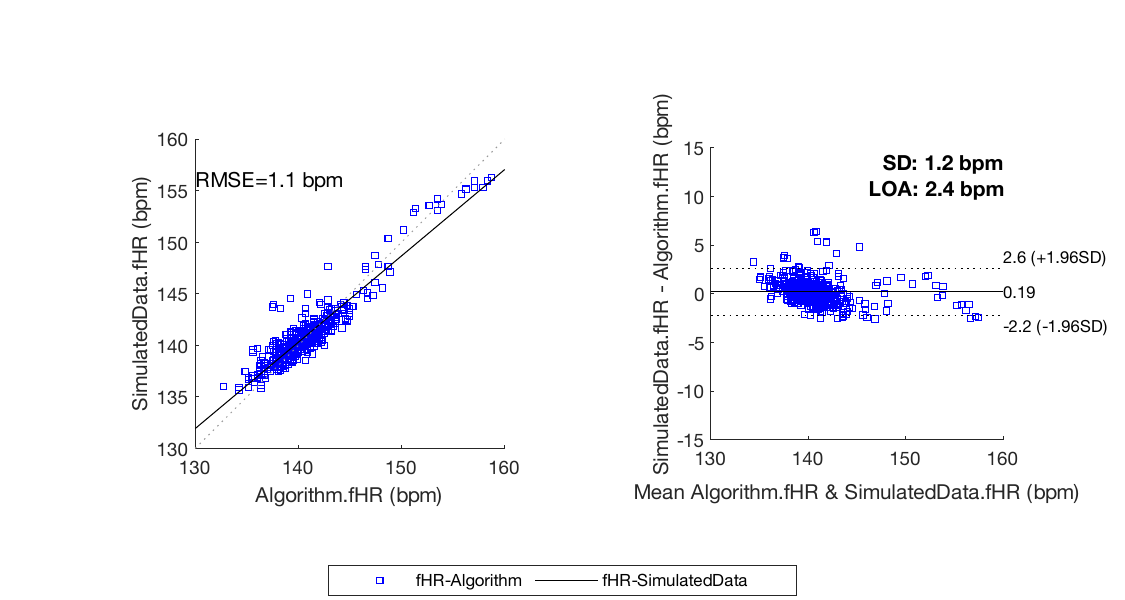}
\caption{Single measurement data (signal of around 8 minutes) for the beat-to-beat cross correlation analysis (Left plot), as well as Bland-Altman plot for the obtained fHR by proposed algorithm and the reference data (Right plot). Here $SNR = -26.7\ dB$, and the p-value is $<0.0001$  for the simulated measurement data.}
\label{fig:Bland-Altman}
\end{figure*}

\begin{figure*}[!t]
  \centering
  \includegraphics[width=\linewidth]{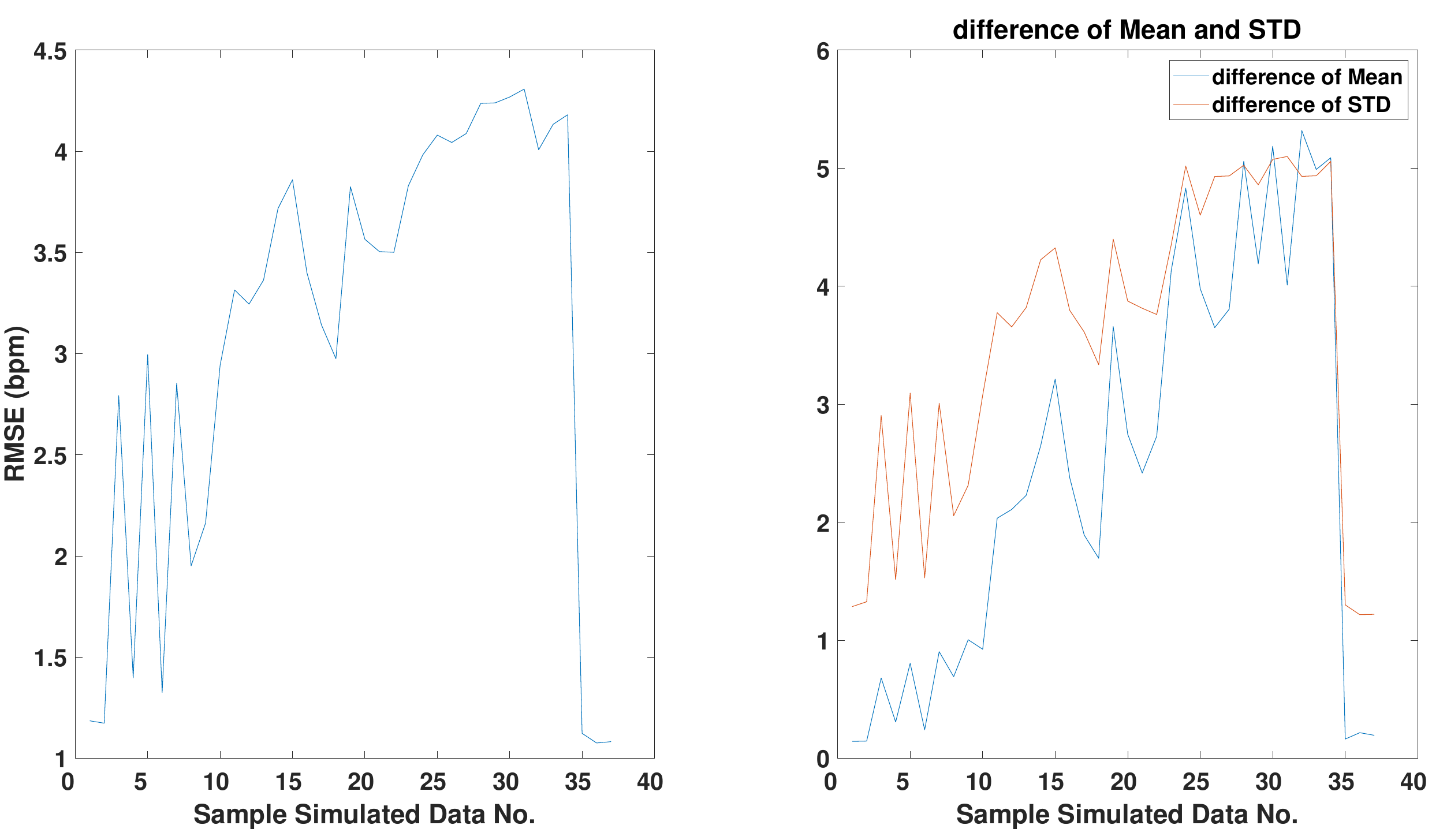}
  \caption{Statistical analysis over the performance of the algorithm is reported. The average p-value is $<0.0008$. (a) root mean square error (RMSE), and (b) mean difference and standard deviation (STD) difference between calculated fHR and baseline for the simulated dataset are plotted.}
  \label{fig:PvalueAndRMSE}
\end{figure*}

In this study, RWEER is calculated until level $j=12$ on 215 abdominal fetal heart sound sections. 85 mother wavelets are investigated to select the most proper one. These mother wavelets are $db1, db2, ..., db45$, $sym1, sym2, ..., sym20$, $coif1, ..., coif5$, and $bior1.1, ..., bior6.8$.
Maximum RWEER for each level is obtained. First row of Fig. \ref{fig:MWI} shows that the highest RWEER happens in level 4. Therefore, level $j = 4$ is chosen. In the next step, RWEER is calculated for $ level = 4$ to find the best mother wavelet. In Fig. \ref{fig:MWI} bottom, the shown boxplot depicts the RWEER distribution for each mother wavelet applied on 215 fHS sections. Also, red-dashed graph illustrates the execution time for each mother wavelet, and black-dash line displays the ratio of mean to standard deviation for RWEER of all 215 sections. Based on this figure, mother wavelet number 74, bior2.2, is chosen to denoise fHS signals. It has the highest mean/std ratio and it also is more repeatable than others while its time execution is low. 

To sum up, in this work, decomposition, denoising and reconstruction of fHS signals were performed by the use of $bior 2.2$ with 4 levels of decomposition. An example of a noisy fHS versus denoised fHS is shown in Fig. \ref{fig:wavthresh}. In Fig. \ref{fig:wavthresh}(a) 1-second of original fHS signal, and \ref{fig:wavthresh}(b) denoised fHS signal are shown using the proposed denoising approach. This will facilitate extracting the S peaks shown in \ref{fig:wavthresh}(c).

\subsection{Systole Extraction}
The detection of systole is essential for fHR extraction (shown in Fig. \ref{fig:S1S1}). In fHS signals, the distance between systole (S) and diastole (D) is much shorter than that in an adult's heartbeat signal. Considering the fact that diastolic duration is longer than systolic duration, D falls at least 100ms after preceding S and at most 200ms before successive S \cite {koutsiana2017WD-FD}. As a result, a reasonable approach for S extraction is to firstly find all the candidates (including S and D) and secondly select S among them.

In first place, two steps including enveloping and finding local maxima are applied to extract all peaks (S and D). Considering Fig. \ref{fig:wavthresh}, the local maximas can be taken into account as potential candidates for S peaks. The procedure of S selection is based on the fact that the normal duration of a fetal beating heart cycle is $430ms$ while the minimum is $375ms$ and the maximum is $545ms$ \cite{koutsiana2017WD-FD}. Therefore, in this study, a thresholding method is chosen to extract S peaks. Selected S-peaks are passed through the next step, which is the fHR calculation. An example of the results of S-selection is shown in Fig. \ref{fig:wavthresh3} . In this figure, the red graph shows the obtained envelope. All candidates are illustrated in red bullet points while selected S-peaks are depicted in green bullet points. 

\subsection{fHR cardiograph visualization}
Knowing that the time elapsed between two successive S in a fHS signal, is a combination of systolic time and diastolic time\cite{kovacs2011fetal}, fHR can be calculated using Eq.\ref{eq:RRHR}. 

\begin{equation}
\label{eq:RRHR}
fHR(bpm)=\frac{60}{T_{{s}{s}}(sec)}
\end{equation}
where, $T_{{s}{s}}$ is the time duration between two S.

For a better projection of fHR when we visualize it in a cardiograph, we smooth the data by looking after possible outliers with high level of volatility in the beat-to-beat analysis, e.g., we remove surrounding noises with high and low frequencies and replace them with a corrected values according to our correction algorithm when applicable. This step is performed for the visualization of the cardiograph and for the sake of avoiding stress level for end users. In the cardio fHR graph we report elaborated fHR datapoints, i.e., in case there is a missing signal datapoints from the device, we keep the nearest fHR measured by the clinical device.

In order to obtain fHR for the graph, different methods such as 'moving median' and 'moving mean' have been used. These methods are used for the sake of correctness of a final cardiograph signal with the objective of less volatile fHR plot which are mainly caused by spontaneous movements and artifacts.

\section{Analysis and validation of results}
The analysis and validation of this study is applied on a personal computer using Matlab R2020a on two set of experiments. Firstly, the public standard simulated fHS database \cite{cesarelli2012simulation} is used for algorithm implementation. This dataset contains 37 simulated fPCG signals. Each signal has a duration of at least 8 minutes, and sampling frequency of 1 kHz. 
In order to evaluate the performance of our algorithm, the analysis of beat-to-beat correlation and Bland-Altman plot \cite{giavarina2015understanding} are performed between obtained fHR and extracted fHR from the reference data (baseline).  

\subsection{Analysis over simulated data}
Firstly we analyze the performance and accuracy of our algorithm for a single measurement signal of about 8 minutes. In Fig. \ref{fig:Bland-Altman} we have calculated the cross-correlation and Bland-Altman statistical analysis which is often used to display of the relationship between two paired variables using the same scale. As it is depicted in the figure, we have obtained the p-value of $<0.0001$. Statistically significance of our result is reported by p-value in the figure: if the p-value falls below the significance level (normally $<0.05$), then the result is statistically significant. Here the confidence interval is chosen to be 95\% and the SNR is reported in the caption of the figure.

Secondly we statistically evaluate our algorithm performance over 37 signals of the simulation sample data, and hence obtain the overall accuracy and correlation of our algorithm. By using a confidence interval of 95\%, the p-value obtained is very promising with an average value of $<0.0008$ for the simulated data (see Fig. \ref{fig:PvalueAndRMSE}). Moreover, for RMSE (bpm) of simulated data an average value of $2.74$ bpm is obtained. In Fig. \ref{fig:PvalueAndRMSE} detailed analysis aggregated by each measurement signal of the simulated dataset is reported.


\begin{equation}
  AverageAccuracy = \frac{1}{N_T}\sum_{i=1}^{K} (\frac{NE_i}{60})
\label{eq:NormalizedAccuracy}
\end{equation}

\begin{figure*}[!t]
  \centering
  \includegraphics[width=\linewidth]{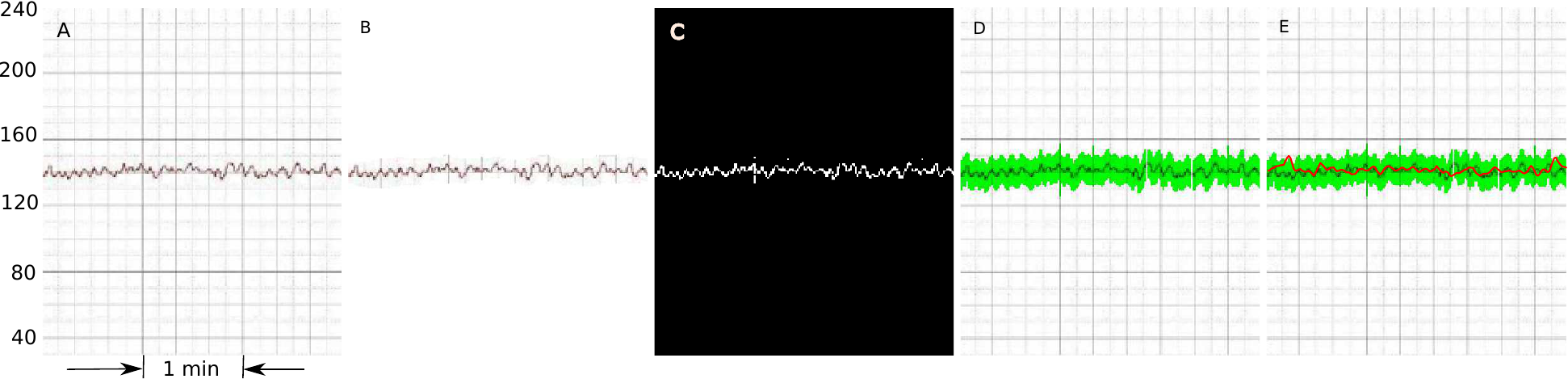}
  \caption{The use of image processing techniques for validation of the proposed algorithm. A) Saved image from clinical device. B,C) fHR graph extraction, D) Confidence interval definition, and E) Adding fHR results obtained from proposed algorithm in red.}
  \label{fig:fHRComparison}
\end{figure*}

\subsection{Analysis over clinical data}

\begin{figure}[!ht]
  \centering
  \includegraphics[width=0.8\linewidth]{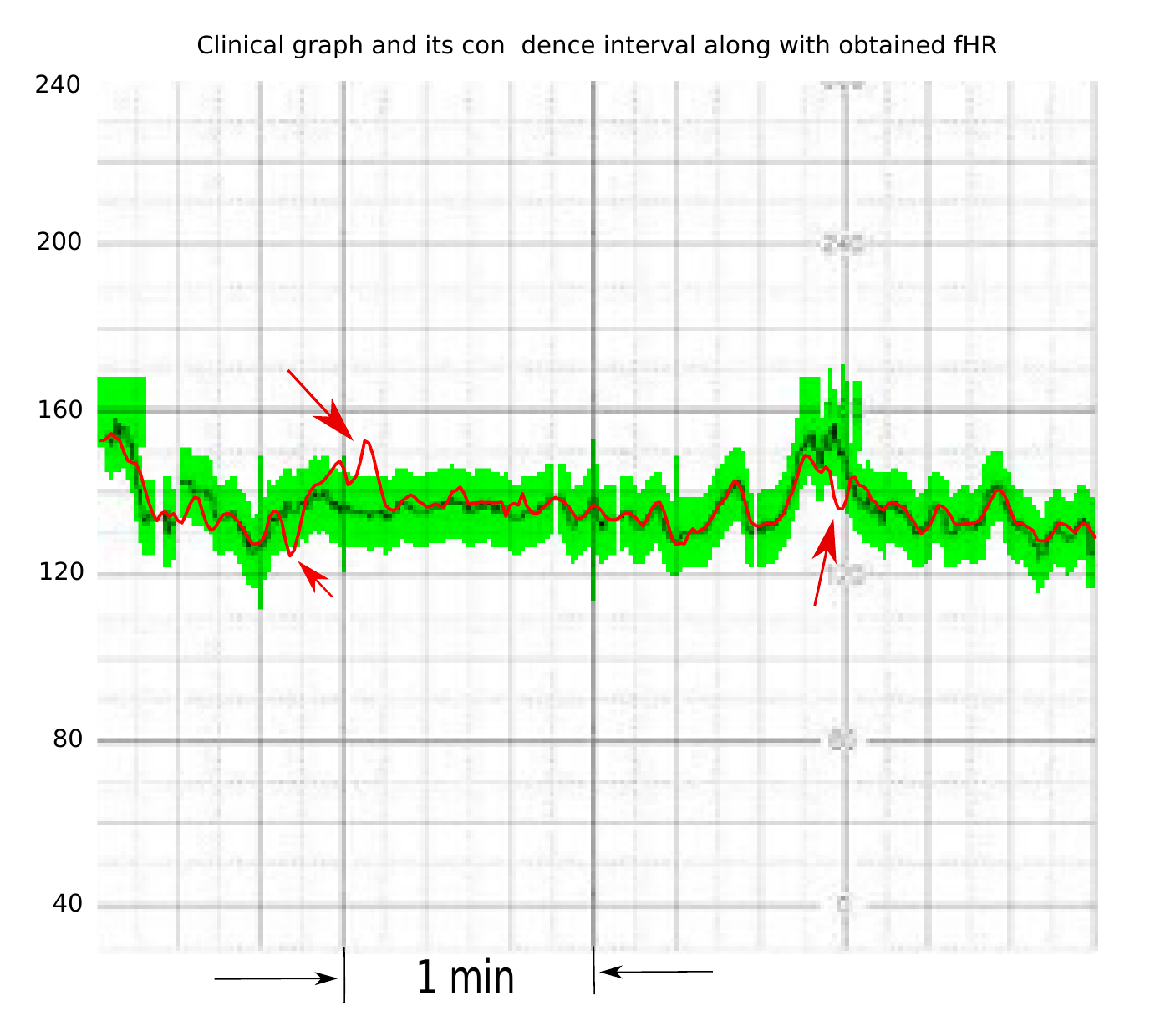}
  \caption{A comparison sample between the algorithm's final graph (in red) and the result of the clinical device (in black). The confidence interval is shown in green. Outliers are shown by red arrows.}
  \label{fig:ComparisonWithArrows}
\end{figure}

Eventually, the algorithm is analyzed and further tuned using the clinical dataset simulatanuosly collected by a portable medical device and the clinical device. For this purpose, the portable Baby Heart Beat fetal Doppler device is used and the fHS signals data are saved by the mobile App. Simultaneously, a clinical device, Bionet model FC-1400 is used to record and the fetal signals with fHR. The ultimate fHR reported by the clinical device is saved in an image format by the device shown in Fig. \ref{fig:ComparisonWithArrows}(A), where y-axis shows fHR value (bpm) while x-axis reflect the measurement timestamp (second). In this image,  the height of each scale square is 10 bpm while its width is 10 second. Using image processing techniques in MATLAB, we extract the clinical fHR from the image (Fig. \ref{fig:ComparisonWithArrows} (B) and (C)). Based on the following medical report \cite{hayes2012accuracy}, calculated fHR by the algorithm could be accepted if its value is $\pm7\ bpm$ than the one reported by a standard clinical device.  Therefore, $\pm 7\ bpm$ is highlighted as a medical acceptance interval in figure \ref{fig:ComparisonWithArrows}(D) (shown in green). Finally, after identifying the confidence interval of our baseline which is the clinical fHR, in Fig.\ref{fig:ComparisonWithArrows}(E) we overlay the obtained fHR with our sophisticated algorithm for the comparison and error rate detection. 

Fig. \ref{fig:ComparisonWithArrows} shows another example in which errors are depicted by red arrows and demonstrates how we detect the errors and optimize the performance of our algorithm.

\begin{figure*}[!b]
\centering
\subfloat[]{\includegraphics[width=\linewidth]{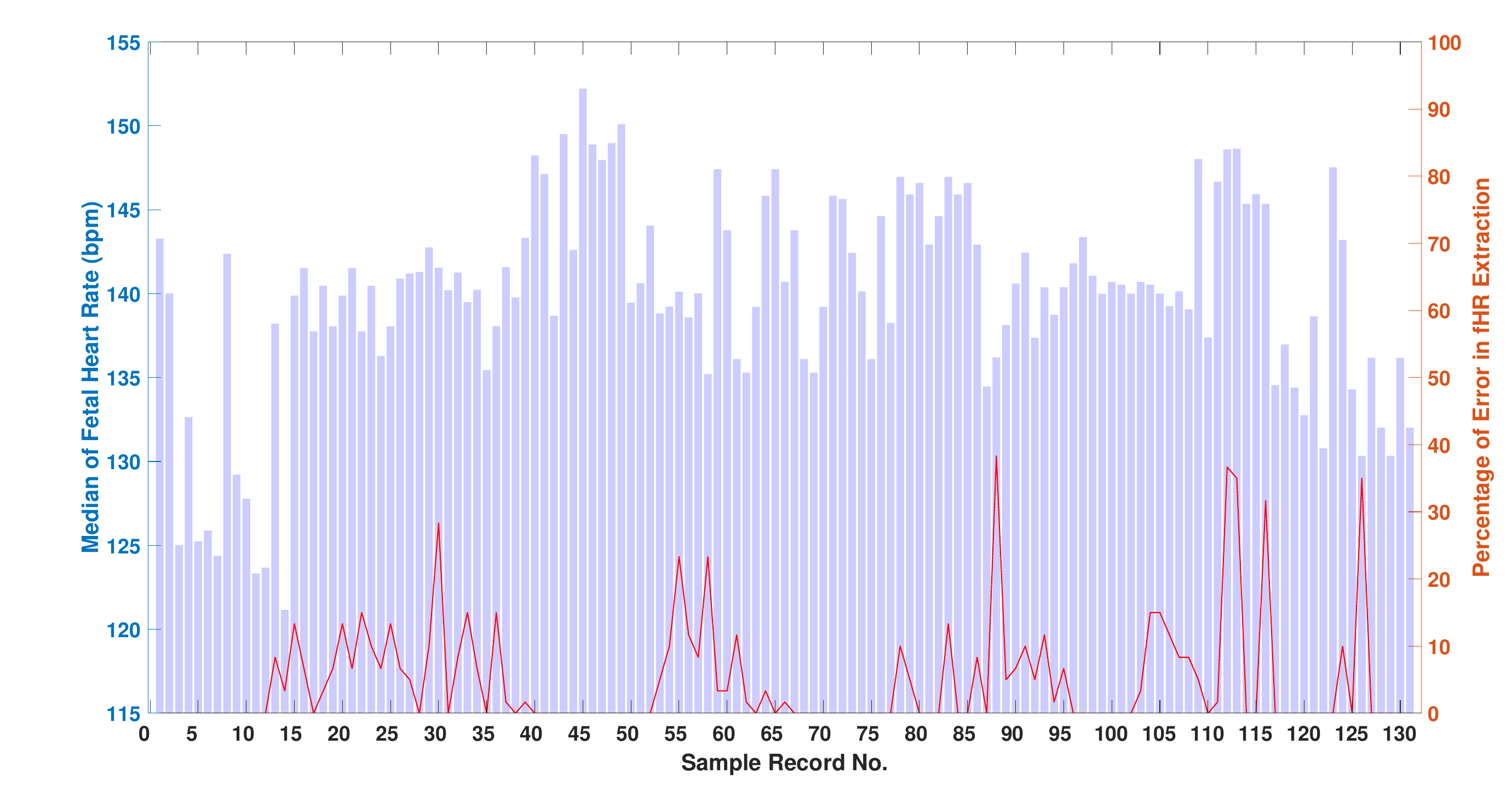}%
\label{fig_first_case}}
\hfil
\subfloat[]{\includegraphics[width=\linewidth]{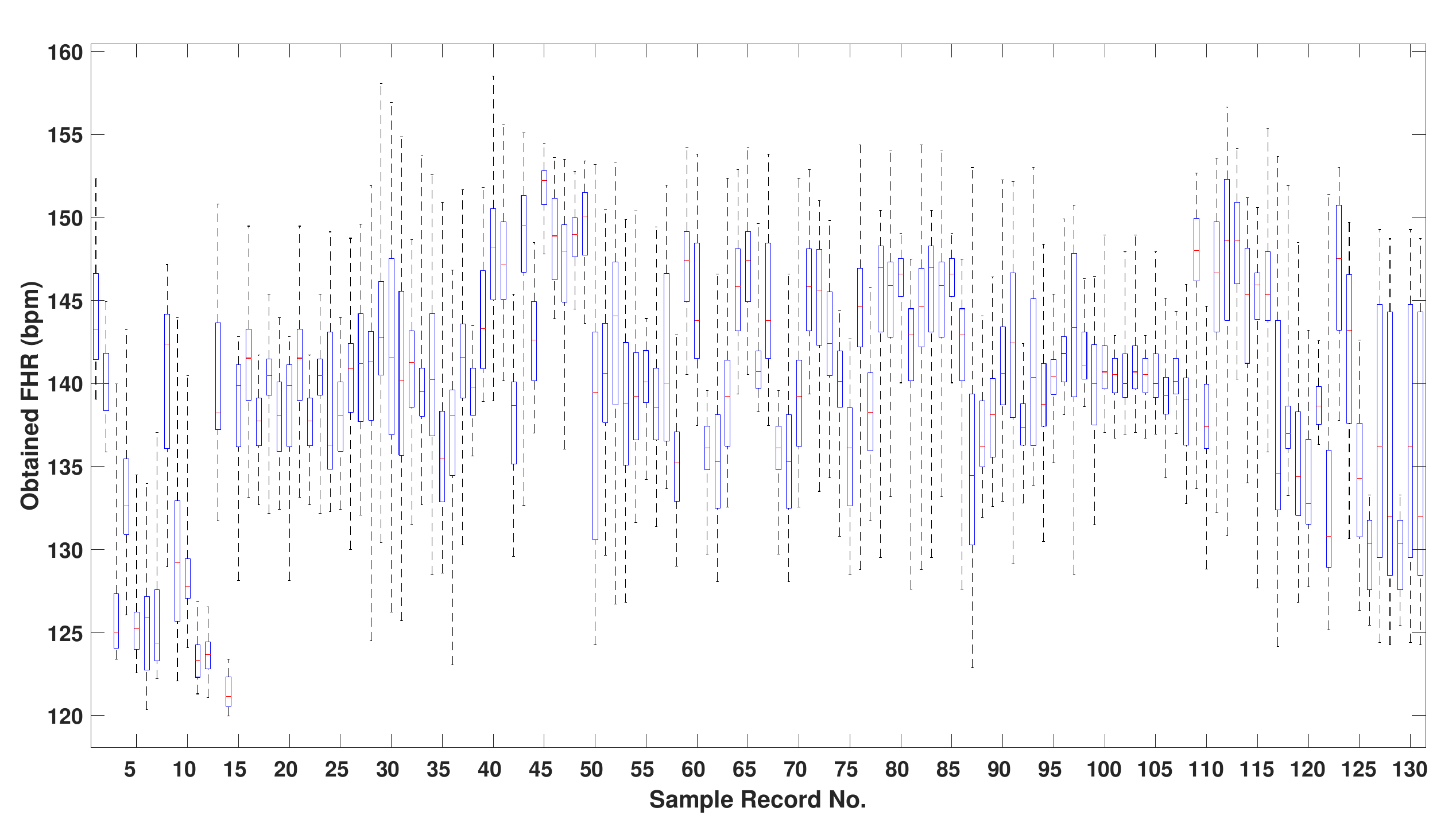}%
\label{fig_second_case}}
\caption{The percentage of error obtained for each data. The orange graph shows the percentage of obtained error. Blue bars illustrate the length of each data in seconds.}
\label{fig:NormalizedAccuracy}
\end{figure*}



Note that along this paper each sample of clinical data is objected to have more than 1-minute of duration. In order to validate the proposed algorithm, an average accuracy is calculated (Eq.\ref{eq:NormalizedAccuracy}). Consider $NE_{i}$ is the number of errors in signal $i$. The length of each signal is 60 seconds, and $N_T$ shows the total length of the used dataset. 
Considering 131 samples of data collected for our investigation, $N_T$ equals 7860 (131 times 60sec of sample data). The total number of reported errors is only 390 which shows an accuracy of 95.03\% and, consequently, an error of 4.96\%. The distribution of the median of calculated fHR as well as the obtained error is shown in Fig. \ref{fig:NormalizedAccuracy}(a). 
Moreover, in order to provide a sense of distribution per measurement we report quantile distribution of the measurement fHR values in the boxplot of Fig. \ref{fig:NormalizedAccuracy}(b). The boxplot reflects the volatility of the fHR and statistical distribution and characteristic of each measurement.

Just to mention that high error rate in some measurements are due to missing datapoints with the clinical device which is often the case. We note that our algorithm can perform much better for those corner cases in compare with the standard clinical device. We believe that by further tuning our algorithm with more clinical data, we would be able to build a groundbreaking approach for fetal monitoring.

\section{Discussion}

\begin{figure*}[!ht]
  \centering
  \includegraphics[width=\linewidth]{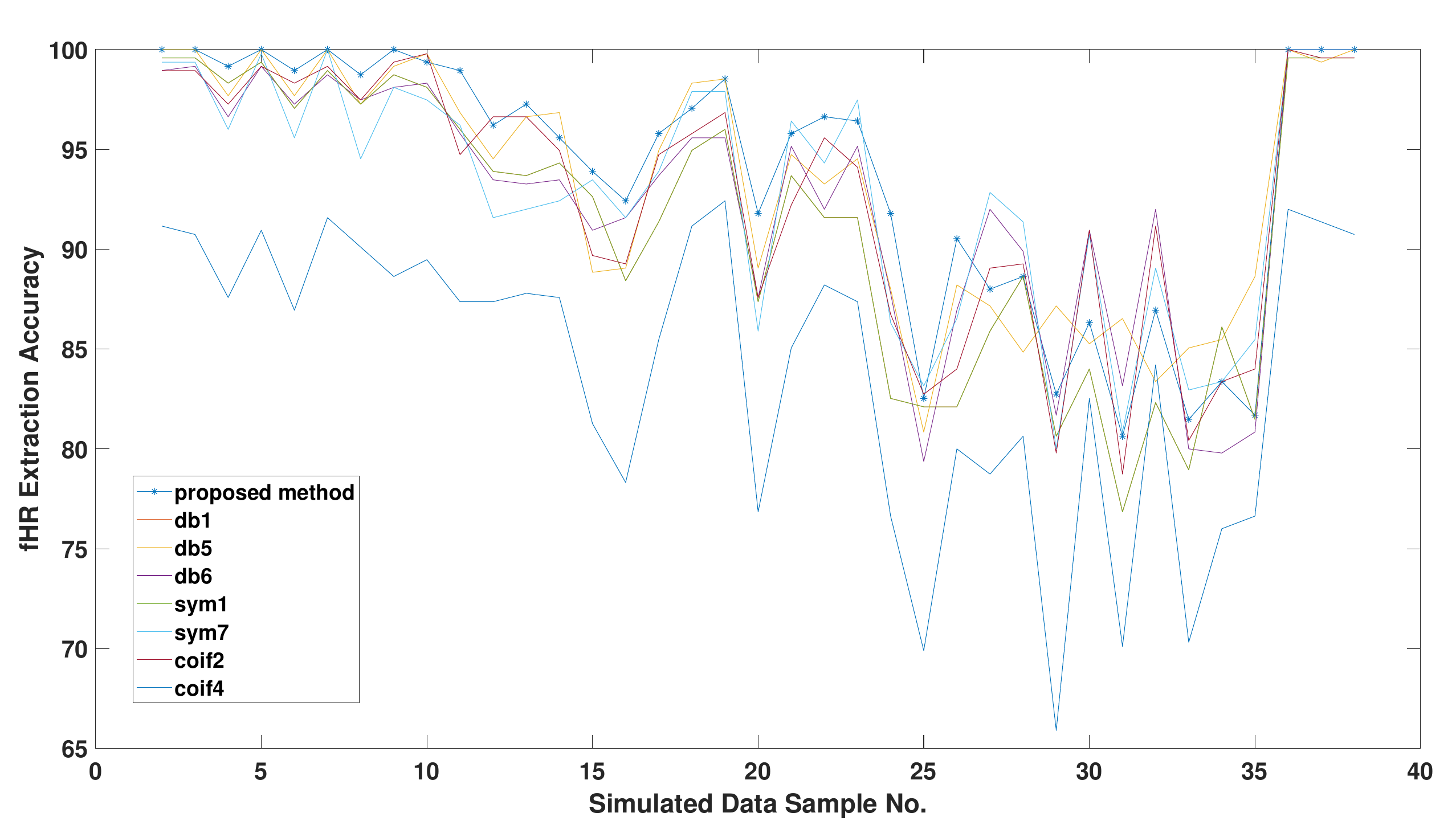}
  \caption{ The results of the comparison of the current work with the works performed by others\cite{tomassini2019wavelet,adithya2017trends}. }
  \label{fig:ComparisonWithOthers}
\end{figure*}

In order to compare the performance of the proposed algorithm versus existing ones, competitive wavelet families including 'db1', 'db5', 'db6', 'sym1' , sym7', coif2' and 'coif4' are selected. To do this, the 37 signals of the simulated dataset were used. The accuracy of the fHR extraction was calculated after using different methods for denoising and presented in Fig. \ref{fig:ComparisonWithOthers}.  The results in this figure illustrate that the applied method provides the best performance in most cases

The fetal heartbeat sound signals are often very noisy. There are several sources of noise related to the mother and fetus as well as the environment. In order to extract the fetal heart rate from fHS signals, denoising is the most important step. The better denoising, the higher accuracy will be achieved in fHR extraction. 
In this paper, a comprehensive study has been performed on fHS denoising using the wavelet technique. We cover many mother wavelets in many different levels of decomposition. Mother wavelet families of Daubechies from 1st order to 45th order, Coiflets from 1st order to 5th order, Symlet from 1st order to 20th order, and Biorthogonal from order 1.1  to the order 6.8 were all explored. All levels from second to 12 were also investigated. In total, we have performed 1020 investigations to cover four mother wavelet families completely. In addition, we captured more than 131 minutes of clinical data from different pregnant women with different BMIs in different weeks of pregnancy. Among captured data, there are mothers with diabetes and high pressure, too. This proves that the proposed algorithm is quite promising in a clinical environment. 
In addition, the embedded software which is provided by the proposed algorithm and is connected to a pocket-size doppler device could help clinicians to reduce unnecessary loads. It also provides a possibility of remote monitoring of high-risk pregnancies with a connected medical device.

\section{Conclusion}
Fetal heartbeat sound signals, generally, have low amplitude and are hidden by high-amplitude noises that may come from the sounds of mother breath, fetal movements and other ambient sounds. In the present paper, an algorithm has been developed for the estimation of fHR from fHS signals. Firstly, a pre-processing task has been done for denoising based on wavelet transformation. Then, a combination of enveloping and finding local maximum is applied for the extraction of systole and diastole peaks. Further, systole peaks have been selected using distance information between S-peaks and D-peaks. Finally, fHR was calculated through the computation of interval times between the S-peaks.

Based on the obtained results and the comparison between those and the ones obtained from simulated/clinical signals, we can conclude that our proposed method is a promising tool for the identification of reliable fHR. In the future, by collecting more hours of annotated data with abnormalities, we would like to address fetal anomaly classification analysis with sophisticated machine learning techniques.

\section*{Acknowledgments}
The authors are grateful to the team of Mehr-e-Madar hospital and dedicated gynecology clinic, for the clinical data acquisition.

\bibliographystyle{IEEEtran}
\addcontentsline{toc}{section}{References}
\bibliography{main}
\end{document}